\documentclass{ws-rv961x669}
\usepackage{ws-rv-van}     % numbered citation/references (default)
\usepackage{ws-rv-thm}     % comment this line when `amsthm / theorem / ntheorem` package is used
\usepackage{subfigure}     % required only when side-by-side / subfigures are used
\makeindex
\begin{document}

\chapter[Spinning protons and gluons in the $\eta'$]
%\\
%Using World Scientific's Review Volume Document Style]
{Spinning protons and gluons in the $\eta'$ 
%
%Using World Scientific's Review Volume\\ Document Style
\label{ra_ch1}}

\author[Steven D. Bass]{Steven D. Bass%\footnote{Author footnote.}
}
%\index[aindx]{Author, F.} % or \aindx{Author, F.}
%\index[aindx]{Author, S.} % or \aindx{Author, S.}
%
\address{
Kitzb\"uhel Centre for Physics, Kitzb\"uhel, Austria \\
Institute of Physics, 
Jagiellonian University in Krakow, Krakow, Poland \\
Steven.Bass@cern.ch
}

\begin{abstract}
The proton spin puzzle has inspired a vast programme of experiments and theoretical work challenging our understanding of QCD and its role in the structure of hadrons.
The proton's internal spin structure is connected to chiral symmetry and,  through gluon degrees of freedom in the flavour singlet channel, 
to the physics of the $\eta'$ meson.
Why do quarks contribute just about one third of the proton's spin? 
Why are $\eta'$ mesons and their interactions so sensitive to gluonic degrees of freedom?
We review the status of these topics and some key observables for forthcoming experiments. 
\end{abstract}

\vspace{3ex}

\centerline{\textsf Dedicated to the Memory of Harald Fritzsch.}

%\markboth{Even Page Header}{Odd Page Header} % Customized running heads

\body

%\tableofcontents

\section{Introduction}

Harald Fritzsch is renowned for his many creative and seminal contributions covering 
the origins and development of QCD, 
electroweak physics, 
ideas with possible unification of fundamental interactions and
cosmology, all driven by a strong physical intuition.
My first meeting with Harald was at the Rencontres de Blois conference in 1994. 
Then followed many stimulating discussions, primarily in Austria and in Munich.
In this contribution I cover a selection of topics 
involving non-perturbative glue in QCD and the transition from current to constituent quarks.
The emphasis will be on the spin structure of the proton and physics of the $\eta'$ meson, 
topics where Harald made initial key contributions.
The present status of this physics is reviewed along with some challenges for 
fresh and forthcoming experiments from high energy deep inelastic scattering 
through to searches for $\eta'$ meson bound states in nuclei. 
The proton's internal spin structure along with the $\eta'$  brings together 
many aspects of QCD 
including confinement dynamics, dynamical chiral symmetry breaking, the QCD axial anomaly and both perturbative and topological properties of gluons as well as fresh ideas about quantum entanglement in QCD.

\section{Gluons and hadrons}

QCD is our theory of strong interactions and the structure of hadrons.
Historically, it developed from the Eightfold Way patterns observed in hadron spectroscopy with wavefunctions described in terms of 
SU(3) flavour, SU(2)  spin and, inside baryons,  
 antisymmetric in a new SU(3) colour label 
plus the parton description of deep inelastic scattering. 
Then came the insight that colour is a dynamical quantum number and the discovery 
of QCD as a non-abelian local gauge theory with coloured gluons as the gauge bosons mediating interactions between quarks and gluons~\cite{Fritzsch:1973pi,Fritzsch:1972jv}.
Asymptotic freedom \cite{Gross:1973id,Politzer:1973fx} 
with the essential role of the non-abelian three gluon vertex 
provided the connection between high energy and low energy phenomenology 
with the realization that the QCD 
coupling $\alpha_s (Q^2)$ decreases logarithmically with increasing momentum transfer $Q^2$, 
with small interaction strength in the ultraviolet and strong interactions in the infrared. 
The glue that binds the proton plays an essential role in its mass and internal spin structure.

Low energy QCD is  characterized by confinement and dynamical chiral symmetry breaking,
DChSB. 
The gluonic confinement potential contributes most of the proton's mass of 938 MeV 
with the rest determined by small quark mass perturbations.
The masses of the proton’s constituent two up quarks and one
down quark are about 2.2 MeV for each up quark and 4.7 MeV for the down quark from
the QCD Lagrangian.
The colour hyperfine one-gluon-exchange potential, OGE, between confined quarks generates the mass splitting between the spin $\frac{1}{2}$ nucleon and its spin $\frac{3}{2}$ 
lowest mass $\Delta$(1232) 
resonance excitation~\cite{Close:1979bt}.
Chiral symmetry is  dynamically broken with formation of a vacuum quark condensate. 
The lightest mass pseudoscalar mesons, the pion and kaon, 
are would-be Nambu-Goldstone bosons associated with DChSB and special 
with their mass squared proportional to the masses of their valence quarks inside, 
$m_P^2 \propto m_q$~\cite{Gasser:1982ap}.
Their isoscalar partners, the $\eta$ and $\eta'$ mesons, 
are sensitive both to Goldstone dynamics and to 
non-perturbative gluon topology in the singlet channel which gives them extra mass and interaction -- see Sect. 4 below.

Hadron masses are  connected to gluonic matrix elements via the trace anomaly in the QCD energy-momentum tensor~\cite{Collins:1976yq,Shifman:1988zk}.
Whereas QCD with massless quarks is classically scale invariant, the proton mass is finite with infrared physics characterized by the infrared scale 
$\Lambda_{\rm qcd} \approx  200$ MeV  
associated with the running QCD coupling $\alpha_s$.
Scale/conformal transformations are associated with 
the scale or dilation current
$d^{\mu} = x_{\nu} \theta^{\mu \nu}$  
with $\theta_{\mu \nu}$ the QCD energy-momentum tensor;
$d_\mu$ satisfies the divergence equation
$\partial_{\mu} d^{\mu} = \theta^{\mu}_{\ \mu}$
with
\begin{equation}
    \theta^{\mu}_{\ \mu} = (1 + \gamma_m) \sum_q m_q {\bar{\psi}_q}\psi_q + \beta (\alpha_s) / 4 \alpha_s  \ G_{\mu \nu}^a G^{\mu \nu}_a. 
\label{eq1}
\end{equation}
This is 
non-vanishing for massless quarks with 
$\beta (\alpha_s) / 4 \alpha_s  \ G_{\mu \nu}^a G^{\mu \nu}_a$ the trace anomaly term.
Here $\gamma_m$ is the quark mass anomalous dimension, 
$
\mu^2  
\frac{d}{d\mu^2} m_q
=
\gamma_m m_q$
with
$\gamma_m
= - \alpha_s / \pi + ... 
$,  
$m_q$ is the renormalized quark mass and $\mu$ is the renormalization scale;
$\beta (\alpha_s)$ is the QCD $\beta$-function
The forward proton matrix element of $\theta_{\mu \nu}$ is 
$    
\langle p, s | \theta_{\mu \nu} | p,s \rangle = p_\mu p_\nu
$
with trace 
$    
\langle p,s | \theta^{\mu}_{\mu} | p,s \rangle = M^2
$
relating the proton mass $M$  squared to the gluonic trace anomaly term~\cite{Jaffe:1989jz}.
(Here $p_\mu$ denotes the proton's momentum vector and $s_\mu$ its spin vector.)
In contrast, pions and kaons would be massless in the chiral limit 
with massless quarks.
Here the gluonic trace anomaly term must vanish. 
Internal binding cancels against individual quark-antiquark terms 
as manifest in, e.g., the Nambu-Jona-Lasino model~\cite{Nambu:1961tp,Klevansky:1992qe} with the massless pions and kaons emerging as Goldstone bosons.
At low resolution the 
three
valence quarks in the proton behave as massive constituent quark quasiparticles through interaction with the vacuum condensate produced by DChSB.

\section{The spin structure of the proton}

The spin structure of the proton has brought many surprises and continues 
to inspire a vast global programme of research to understand 
QCD spin dynamics. 
Key experiments include inclusive and semi-inclusive 
polarized deep inelastic scattering and high energy polarized proton-proton collisions.

One finds that 
just about 30\% of the proton's spin is carried by the spin of the quarks inside.
This result is deduced from measurements of the 
proton's $g_1(x,Q^2)$ 
deep inelastic spin structure function; 
$Q^2$ denotes minus the four-momentum transfer squared in the deep inelastic process and $x$ is the Bjorken variable.
One evolves all data points with $Q^2 > 1$ GeV$^2$ using next-to-leading order, NLO, 
QCD evolution to the same $Q^2$ and then takes the first moment:
\begin{eqnarray}
\int_0^1 dx \ g_1^p (x,Q^2)
&=&
\Biggl( {1 \over 12} g_A^{(3)} + {1 \over 36} g_A^{(8)} \Biggr)
\Bigl\{1 + \sum_{\ell\geq 1} c_{{\rm NS} \ell\,}
\alpha_s^{\ell}(Q)\Bigr\}
\nonumber \\
& &
+ {1 \over 9} g_A^{(0)}|_{\rm inv}
\Bigl\{1 + \sum_{\ell\geq 1} c_{{\rm S} \ell\,}
\alpha_s^{\ell}(Q)\Bigr\}  +  {\cal O}({1 \over Q^2})
 + \ \beta_{\infty}
.
\label{eq2}
\end{eqnarray}
Here $g_A^{(3)}$, $g_A^{(8)}$ and $g_A^{(0)}|_{\rm inv}$ are the
isovector, SU(3) octet and scale-invariant  flavour-singlet
axial-charges respectively.
The flavour non-singlet $c_{{\rm NS} \ell}$
and singlet $c_{{\rm S} \ell}$ Wilson coefficients
are calculable in $\ell$-loop perturbative QCD.
These perturbative QCD coefficients have been calculated
to $O(\alpha_s^3)$ precision \cite{Larin:1997qq,Blumlein:2022gpp}.
For $\alpha_s = 0.3$ typical of the deep inelastic experiments one finds
$\Bigl\{1 + \sum_{\ell= 1}^3 c_{{\rm NS} \ell\,}
\alpha_s^{\ell}(Q)\Bigr\}
=
0.85
$
and
$\Bigl\{1  + \sum_{\ell = 1}^3 c_{{\rm S} \ell\,}
\alpha_s^{\ell}(Q)\Bigr\}
=
0.96
$.
The term $\beta_{\infty}$
represents a possible leading-twist subtraction constant from
the circle at infinity when one closes the contour in the
complex plane in the dispersion relation
\cite{Bass:2004xa}.
The subtraction constant affects just the first moment and corresponds to a possible contribution at Bjorken $x$ equal to zero.

In terms of the flavour-dependent axial-charges
\begin{equation}
2M s_{\mu} \Delta q =
\langle p,s |
{\overline q} \gamma_{\mu} \gamma_5 q
| p,s \rangle
\label{eq3}
\end{equation}
the isovector, octet and singlet axial charges are
\begin{eqnarray}
g_A^{(3)} &=& \Delta u - \Delta d
\nonumber \\
g_A^{(8)} &=& \Delta u + \Delta d - 2 \Delta s
\nonumber \\
g_A^{(0)}|_{\rm inv}/E(\alpha_s)
\equiv
g_A^{(0)}
&=& \Delta u + \Delta d + \Delta s
.
\label{eq4}
\end{eqnarray}
The singlet axial-charge comes with a two-loop anomalous dimension and is often quoted as the 
value evolved to infinity using 3 flavour QCD evolution, coinciding with the scale invariant form $g_A^{(0)}|_{\rm inv}$ with the renormalization scale dependence parametrized by the factor $E(\alpha_s)$ factored out.

In the parton model the quantities $\Delta q$ are interpreted 
(before gluonic effects in the flavour-singlet channel) as the fraction of the proton's spin carried by quarks and antiquarks of flavour $q$.
The isovector axial-charge is measured independently in neutron
$\beta$-decays
($g_A^{(3)} = 1.275 \pm 0.001$~\cite{Workman:2022ynf})
and the octet axial charge is commonly taken to be the value 
extracted from hyperon $\beta$-decays assuming a two-parameter 
SU(3) fit ($g_A^{(8)} = 0.58 \pm 0.03$ \cite{Close:1993mv}).
The SU(3) symmetry assumption here
may be strongly broken, e.g., by pion cloud effects -- see below --
and the error on $g_A^{(8)}$ 
could really be as large as 25\%~\cite{Jaffe:1989jz}.

With this input, 
the polarized deep inelastic scattering experiments are interpreted
in terms of a small value for the flavour-singlet axial-charge.
Using the SU(3) value for $g_A^{(8)}$ and assuming no leading twist 
subtraction constant COMPASS found~\cite{COMPASS:2016jwv}
\begin{equation}
g_A^{(0)}|_{\rm pDIS, Q^2=3 {\rm GeV}^2}
=
0.32 \pm 0.02 ({\rm stat.}) \pm 0.04 ({\rm syst.}) \pm 0.05 ({\rm evol.})
\label{eq5}
\end{equation}
or
$
g_A^{(0)}|_{\rm pDIS, Q^2 \rightarrow \infty} 
=
0.31 \pm 0.06
$
taking into account QCD renormalization group evolution.
(This deep inelastic quantity misses 
 any contribution to $g_A^{(0)}|_{\rm inv}$ 
 from a possible delta function at $x=0$).
When combined with $g_A^{(8)} = 0.58 \pm 0.03$,
the value of $g_A^{(0)}|_{\rm pDIS}$ in 
Eq.~(\ref{eq5})
corresponds to a negative strange-quark polarization
$
\Delta s_{Q^2 \rightarrow \infty}
=
\frac{1}{3}
(g_A^{(0)}|_{\rm pDIS, Q^2 \rightarrow \infty} - g_A^{(8)})
=
- 0.09 \pm 0.02
$
-- that is,
polarized in the opposite direction to the spin of the proton.
With this $\Delta s$, the following values for the
up and down quark polarizations are obtained:
$
\Delta u_{Q^2 \rightarrow \infty}
=
0.84 \pm 0.02$
and 
$
\Delta d_{Q^2 \rightarrow \infty}
=
-0.44 \pm 0.02
$.
As a consistency check, 
the Bjorken sum-rule relates the difference in the first moments of $g_1$ for proton and neutron targets to the isovector $g_A^{(3)}$ 
with extracted value 
$1.29 \pm 0.05 \pm 0.10$~\cite{COMPASS:2016jwv}. 
This value agrees well with the number from neutron $\beta-$decays, 
so here the theory is working as it should.

The value in Eq.~(\ref{eq5}) 
compares with the estimates of about 0.6 from 
the simplest relativistic quark models like the MIT Bag, which associates 
the extra 40\% 
with quark orbital angular momentum. The value 0.6 is also the 
value one would expect by taking 
the octet axial charge if extracted assuming good SU(3) flavour symmetry in the hyperon 
$\beta$-decays and assuming $\Delta s=0$.
The initial EMC measurement at CERN 
inspired considerable surprise with a first value of $g_A^{(0)}$ consistent with zero~\cite{EuropeanMuon:1987isl}.

Looking in the data the small value of $g_A^{(0)}$ 
measured in the experiments 
is associated with a collapse in the isoscalar deuteron spin 
structure function 
to something close to zero in the low $x$ region between 0.004 and 0.05. 
The convergence of the moment 
integrals is shown in Fig.~\ref{fig:g1int}.  Here ``Bjorken''  denotes the theoretical expectation for the isovector sum-rule 
$\int_0^1 dx g_1^{(p-n)}$ 
and ``Ellis-Jaffe''  denotes the expectation for the isosinglet combination
$\int_0^1 dx g_1^{(p+n)}$
if the 
strange term $\Delta s$ 
were zero and if good SU(3) symmetry were working with the nucleon's octet axial charge. 
One observes that the proton spin puzzle is associated with this 
collapse in the isosinglet structure function for 
$x \lesssim 0.05$ 
which needs to be understood.
This is in contrast  with the isovector spin structure function which continues to rise with decreasing $x$  
and with unpolarized deep inelastic scattering, where the low $x$ structure function is dominated by large isoscalar gluonic pomeron exchanges~\cite{Landshoff:1994up}.

Key issues are the  interpretation of the flavour-singlet axial-charge and 
possible SU(3) breaking in the octet term used to extract it.
\begin{figure}[tpb]
\centering
\includegraphics[width=0.82\textwidth]{g1_int.eps}
\caption{\small
Convergence of the first moment integral of $g_1$ as a
function of the lower integration limit $x_{\rm min}$ 
for the Bjorken integral 
(isospin non-singlet) and 
the Ellis--Jaffe integral (iso-singlet)
from the COMPASS proton and deuteron data evolved to  
$Q^2=3$~GeV$^2$.
The arrows indicate the theoretical expectations.
Error bars are statistical errors only. Figure from Ref.~\cite{Aidala:2012mv}.
}
\label{fig:g1int}
\end{figure}
Theoretical QCD analysis based on the axial anomaly
leads to the formula
\begin{equation}
g_A^{(0)}
=
\biggl(
\sum_q \Delta q - 3 {\alpha_s \over 2 \pi} \Delta g \biggr)_{\rm partons}
+ {\cal C}_{\infty}
\label{eq8}
\end{equation}
-- see Refs.~\cite{Altarelli:1988nr,Efremov:1988zh,Carlitz:1988ab,Bass:1991yx,Bass:2004xa}.
Here $\Delta g_{\rm partons}$ is the amount of spin carried by polarized
gluon partons in the polarized proton
with
$\alpha_s \Delta g \sim {\tt constant}$ 
as $Q^2 \rightarrow \infty$
\cite{Altarelli:1988nr,Efremov:1988zh} and the growth in gluon polarization at large $Q^2$ compensated by similar growth with opposite sign in the gluon orbital angular momentum;
$\Delta q_{\rm partons}$ measures the spin carried by quarks and
anti-quarks carrying ``soft'' transverse momentum
$k_t^2 \sim {\cal O}(P^2, m^2)$ 
where $P^2$ is a typical gluon virtuality in the nucleon
and $m$ is the light quark mass.
The polarized gluon term is associated with events in polarized
deep inelastic scattering where the hard photon strikes a
quark or anti-quark generated from photon-gluon fusion and
carrying $k_t^2 \sim Q^2$ \cite{Carlitz:1988ab}.
It is associated with the QCD axial anomaly in perturbative QCD.

${\cal C}_{\infty}$ denotes a potential non-perturbative gluon
topological contribution
with support only at Bjorken $x=0$ \cite{Bass:2004xa}.
It is associated with a possible subtraction constant in the
dispersion relation for $g_1$ and, if finite, it would be associated with 
a $J=1$ Regge fixed pole with non-polynomial residue. 
If non-zero it would mean that
$\lim_{\epsilon \rightarrow 0} \int_{\epsilon}^1 dx g_1$ 
will measure
the difference of
the singlet axial-charge and the subtraction constant contribution;
that is, polarized deep inelastic scattering measures the combination
$g_A^{(0)}|_{\rm pDIS} = g_A^{(0)} - {\cal C}_{\infty}$.
Any finite ${\cal C}_\infty$ term  would show up as a difference \cite{Bass:1997zz} in the value of $\Delta s$ 
extracted from polarized deep inelastic scattering and from 
elastic neutrino proton scattering,   which measures the combination 
$(\Delta u - \Delta d - \Delta s)_{Q^2 \to \infty}$ up to small heavy quark radiative corrections, evaluated to NLO accuracy in 
Ref.~\cite{Bass:2002mv}\ ,
and with any ${\cal C}_\infty$ term included in the full axial-current matrix elements. 
Leading twist subtraction constant corrections proportional to 
$\delta (x)$ 
are known in unpolarized deep inelastic scattering through the Schwinger term sum-rule for the longitudinal structure function $F_L(x,Q^2)$~\cite{Broadhurst:1973fr}.

How should we understand the parton model formula, Eq.~(\ref{eq8}), 
in the context of the operator product expansion description of  polarized deep inelastic scattering?
The connection is subtle.
There is no spin-one, local, gauge invariant gluonic operator with the 
quantum numbers of the flavour-singlet axial vector current.
However, gluonic input does enter through the QCD axial anomaly. 
The gauge invariantly renormalized flavour-singlet axial-vector current
$J_{\mu 5} =
\sum_q {\bar \psi}_q  \gamma_{\mu} \gamma_5  \psi_q$ 
satisfies the anomalous divergence equation
\begin{equation}
\partial^{\mu} J_{\mu 5} = \sum_q 2 m_q {\bar \psi}_q i \gamma_5 \psi_q + 3 \frac{\alpha_s}{4 \pi} G_{\mu \nu}^a{\tilde G}_a^{\mu \nu} .
\label{eq9}
\end{equation}
Here 
$G_{\mu \nu}^a$ is the gluon field tensor and 
${\tilde G}_{\mu \nu}^a$ the corresponding dual tensor.
The gauge invariant  current can be written as the sum 
$
    J_{\mu 5} = J_{\mu  5}^{\rm con} + 2f K_{\mu} 
$ 
with
$\partial^{\mu} K_{\mu} = \frac{\alpha_s}{8 \pi} G.{\tilde G}$ 
being a total divergence
involving the gauge dependent Chern-Simons current
$
K^{\mu} = \frac{g^2}{32\pi^2} 
\epsilon^{\mu \alpha \beta \gamma} A_{\alpha}^a
(G_{\beta \gamma}^a - \frac{1}{3} g c^{abc} A_{\beta}^b A_{\gamma}^c)
$
with 
$\alpha_s = g^2/4 \pi$
and
$\partial^{\mu} J_{\mu 5}^{\rm con} = \sum_q 2 m_q {\bar \psi}_q i \gamma_5 \psi_q$;
$f=3$ is the number of active light flavours.
The partially conserved axial-vector current 
$J_{\mu 5}^{\rm con}$
is not gauge invariant.

The QCD parton model is formulated in the light-front 
$A_+=0$ gauge with the leading twist term measured in the forward matrix element of 
the ``+ component'' of the 
flavour-singlet current, viz. $J_{+5}$. 
While $K_\mu$ is gauge dependent, the forward matrix elements of 
$K_+$ 
are actually invariant 
under residual gauge degrees of freedom in the light-front gauge $A_+=0$. 
Subject to these constraints, it is coincident with the ``+ component'' of the gauge invariant 
gluon spin operator 
(up to a possible surface term)
~\cite{Jaffe:1995an,Manohar:1990kr}. 
In more general gauges, one runs into the issue of invariance under large gauge transformations \cite{Jaffe:1989jz} 
which connect to non-perturbative gluon topology beyond perturbation theory and involve 
shuffling a non-local zero mode around between ``quark'' and ``gluonic'' terms \cite{Bass:1997zz}. 
If the net zero mode term is finite, 
then it corresponds to a $x=0$ term in the spin dependent 
parton distributions and a  subtraction at infinity in the $g_1$ dispersion relation~\cite{Bass:2004xa}. 

One can re-label the quark and gluon contributions to the spin structure functions through different choices of factorization 
``scheme'' (or jet definitions) and include all 
the glue terms in Eq.~(\ref{eq8}) 
as  absorbed in the  ``quark'' piece - e.g. in the $\overline{\rm MS}$ scheme. 
For polarized deep inelastic scattering 
the decomposition in Eq.~(\ref{eq8}) is 
``physical'' in the  sense that the different terms correspond to measurable different jet processes with different $k_t$.
The polarized gluon contribution 
$\Delta g$
enters $g_1$ at small $x$ after convolution 
with 
the gluon coefficient 
or hard part of the polarized $\gamma^* g$ 
cross section $C^g$, 
viz.
$\sim \int_x^1 dx \Delta g (z) C^g(\frac{x}{z}, \alpha_s)$. 
Within the parton picture the axial anomaly term corresponds to a term 
$-\frac{\alpha_s}{\pi} (1-z)$ in $C^g$ 
with $z$ the Bjorken variable for the $\gamma^*g$ collision~\cite{Bass:1992ti}.

There is evidence in the data from polarized proton-proton collisions at RHIC for modest polarized gluon spin in the proton.  
Semi-inclusive measurements of fast kaon production in 
polarized deep inelastic scattering reveal no strong evidence for  polarized strangeness~\cite{Aidala:2012mv}.

Specifically, 
with gluon polarization a NLO global fit to spin data including 
from RHIC polarized proton-proton collisions
gives 
$\int_{0.05}^1 dx \Delta g(x) \approx 0.2 \pm 0.05$ and
$|\int_{0.001}^1 dx \Delta g (x) | \lesssim 0.8$, each at $Q^2=10$ GeV$^2$~\cite{deFlorian:2014yva}. 
This interesting result is however not sufficient 
to alone 
resolve the spin puzzle 
through the polarized gluon term 
$-3\frac{\alpha_s}{2\pi}\Delta g$
(which would need a 
value $\Delta g \sim 2$ with $\alpha_s \approx 0.3$ 
assuming with good SU(3) in the determination of $g_A^{(8)}$) 
though it is consistent with theoretical model estimates 
$|\Delta g (m_c^2) |\lesssim 0.3$~\cite{Bass:2011zn} and $\Delta g (1 {\rm GeV}^2) \approx 0.5$~\cite{Brodsky:1989db}.
Other more recent fits 
to $\Delta g(x,Q^2)$ 
including 
sensitivity to the overall sign of $\Delta g$ 
are reported in 
Refs.~\cite{deFlorian:2019egz,Zhou:2022wzm}.
Improved constraints will come from future data with the Electron Ion Collider, EIC.

If $\Delta g$ is too small to explain the small value of $g_A^{(0)}$, 
then what about  possible SU(3) breaking in the nucleon's axial charges?
SU(3) breaking is induced through virtual pion cloud effects.
QCD inspired models 
include both 
the colour hyperfine OGE potential 
responsible for the nucleon-$\Delta$ mass splitting~\cite{Close:1979bt} and the pion cloud induced by dynamical chiral symmetry breaking. 
Taken alone 
(before pion effects) 
OGE 
gives the SU(3) $F/D$  ratio and 0.58 value for the nucleon's octet axial charge extracted assuming  good SU(3) symmetry. 
When the pion cloud is also included along with small kaon loop corrections, 
re-evaluation of the nucleon's 
axial-charges
in the Cloudy Bag model led to the value
$g_A^{(8)} = 0.46 \pm 0.05$~\cite{Bass:2009ed}. 
In these calculations 
$g_A^{(3)}$ retains  its physical
value.
If one instead uses this
new octet number in the analysis of polarized deep inelastic scattering, 
then the corresponding 
values extracted from the experiments become 
$g_A^{(0)}|_{\rm pDIS, Q^2 \to \infty}$ 
$0.33 \pm 0.06$ with
$
\Delta s 
\sim -0.04 \pm 0.03 
$.
A recent 
joint fit to spin dependent parton distributions and fragmentation functions from 
inclusive and semi-inclusive deep inelastic scattering
as well as inclusive $e^+e^-$ data 
gives the  octet term peaked close to 0.5~\cite{Ethier:2017zbq}, 
with value $0.46 \pm 0.21$, 
which is close to the Cloudy Bag  preferred value 
though with large uncertainty.
Recent lattice calculations~\cite{Alexandrou:2017oeh,Liang:2018pis}  
give values of 
$g_A^{(0)} \approx 0.40 \pm 0.04$ 
with
$\Delta s$ close to -0.04.

Summarizing the present status of this phenomenology, the OGE potential plus pion cloud effects 
together with the modest polarized glue suggested by theory and by the RHIC spin experiments 
are sufficient to resolve the small value of $g_A^{(0)}$ 
within the present experimental and theoretical errors.
New data will come from the future EIC  
which will push the measurements to smaller Bjorken $x$, 
down to $x \sim 10^{-4}$,  
and with improved precision.
One issue is the behaviour of $g_1$ at very small $x$, 
below the $x_{\rm min}=0.004$ achieved by COMPASS,  
where perturbative QCD resummation calculations~\cite{Adamiak:2023okq} 
predict more divergent 
small $x$ behaviour 
for the singlet part of $g_1$. 
Predictions in EIC kinematics are given in Ref.~\cite{Adamiak:2021ppq}. 

In parallel to high energy spin experiments, a  precise measurement of elastic neutrino proton scattering would 
be valuable as a  complementary probe of $\Delta s$ and to test gluon topology ideas. 
Here, the present 
most accurate measurement is from KamLAND~\cite{KamLAND:2022ptk}, 
$\Delta s = -0.14^{+0.25}_{-0.26}$.

Confinement induces a parton transverse momentum 
scale and hence finite quark and gluon orbital angular momentum in the proton. 
The quark and gluon total angular momentum are related to terms in 
the corresponding QCD energy-momentum tensor allowing one  to deduce information about their asymptotic behaviour 
where the ratio of quark to gluon contributions becomes 16:3$f$  
with $f$ the number of quark flavours  \cite{Ji:1996ek}.
Beyond this observation,  
specific total and orbital angular momentum definitions are not unique
\cite{Jaffe:1989jz,Ji:1996ek} 
with different definitions being more suitable to the interpretation of different spin processes. 
There is a vigorous  experimental programme to measure $k_t$ 
dependent processes and spin-momentum correlations 
involving transverse-momentum dependent 
and 
non-forward generalized parton distributions, TMDs and GPDs,
to map out the 
tomography of the proton -- for a review see Ref.~\cite{Aidala:2012mv}.

Away from the forward direction the matrix elements of $K_\mu$ are not gauge invariant at all.
This means that for non-forward, 
transverse momentum dependent processes 
like polarized deeply virtual Compton scattering ~\cite{Bhattacharya:2022xxw,Bhattacharya:2023wvy}
one should use the $\overline{\rm MS}$
scheme with the 
full gauge invariant axial vector current including the anomaly included in the spin dependent ``quark'' distribution. 
The decomposition in
Eq.~(\ref{eq8}) 
with the polarized gluon term 
linked to the Chern-Simons current only makes sense in the forward direction~\cite{Bass:2001dg,Bass:2004xa}.

QCD quantum entanglement effects might be important in semi-inclusive 
processes with spin and 
transverse momentum dependence~\cite{Aidala:2021pvc,Kharzeev:2021nzh}. 
In hadron scattering processes 
colour Wilson lines might overlap between 
the incoming (or outgoing) hadrons leading to novel factorization breaking effects~\cite{Aidala:2021pvc}.
Improved experimental precision will be necessary to test these QCD entanglement ideas.

Beyond the parton model picture encompassed in Eq.~(\ref{eq8}) 
further attempts to 
understand the role of gluonic spin degrees of freedom in the 
transition from current to constituent quarks have been explored in 
Refs.~\cite{Fritzsch:1988ht,Fritzsch:1989ty,Jaffe:1989jz,Bass:2004xa,Forte:1990xb,Shore:1991dv,Brodsky:1989db,Tarasov:2020cwl,Tarasov:2021yll}.
The singlet axial charge $g_A^{(0)}$ 
can also be related to the proton 
matrix element of the topological charge density.
Consider first the non-forward matrix element 
\begin{equation}
    \langle p,s | J_{\mu 5} | p', s' \rangle 
    =
    2 M {\tilde s}_\mu G_A (l^2) + l_\mu l.{\tilde s} G_P(l^2)
\label{eq10}
\end{equation}
where $l_\mu = (p-p')_\mu$ and
${\tilde s}_{\mu}
= {\overline u}_{(p,s)} \gamma_{\mu} \gamma_5 u_{(p',s')} / 2M $.
Since the $\eta'$ meson has finite mass even in the chiral limit, it follows that there 
is no massless pole in $G_P(l^2)$ even in the chiral limit. 
Next define the chiralities 
$q.{\tilde s} \chi^q(l^2) = 
\langle p, s| 
    2m_q i \bar{q} \gamma_5 q | p', s' \rangle
$
for each flavour $q$ 
and
the gluonic term 
$q.{\tilde s} \chi^g(l^2) =
\langle p, s| 
    \ \frac{\alpha_s}{4 \pi} G_{\mu \nu}^a{\tilde G}_a^{\mu \nu}  \ | p', s' \rangle
$.
Contracting Eq.~(\ref{eq10}) with $l^\mu$ and taking the limit $l^2 \to 0$ gives
\begin{equation}
2M g_A^{(0)} = 
   2M G_A(0) = 
    \sum_q \chi^q(0) + 2 f \chi^g (0).
\label{eq11}
\end{equation}
With exactly massless quarks 
this involves just the 
$\frac{\alpha_s}{4 \pi} G_{\mu \nu}^a{\tilde G}_a^{\mu \nu}$ topological charge density.
However, 
with even very small quark masses, the quark and gluon terms here come 
proportional to the ratio of the light up and down quark masses~\cite{Veneziano:1989ei}.
This quark mass  dependence cancels in each of the isovector, octet and full singlet axial charges when one sums over all the contributing terms.
If the ``gluon'' term is taken alone, $\chi^g(0)$ is sensitive to large isospin violation.
If one generalizes this discussion to polarized deep inelastic scattering from a polarized real photon target, then the corresponding 
``gluonic term'' 
$\chi^g(0)|_{\gamma}$
for a polarized photon target evaluates~\cite{Bass:1991sg} to 
$2 \pi / \alpha_s \approx +30$ if we take the ratio of 
light quark masses 
$m_u/m_d = \frac{1}{2}$!
The infrared quark mass ratio dependence here contrasts with the $-\frac{\alpha_s}{2\pi}\Delta g$ term in Eq.~(\ref{eq8}) which corresponds to the hard perturbative QCD process of two-quark-jet production at large $k_t^2 \sim Q^2$.

One can also write a flavour singlet Goldberger-Treiman relation connecting $g_A^{(0)}$ and the $\eta'$-nucleon coupling constant~\cite{Shore:1991dv}
\begin{equation}
    2M g_A^{(0)} = F g_{\eta' NN}  
    +
    \frac{1}{6} F^2 m_{\eta'}^2 g_{GNN}.
\label{eq12}
\end{equation}
Here $g_{\eta' NN}$ is the $\eta'$ nucleon coupling constant, 
$g_{GNN}$ 
denotes the proper vertex for coupling of the gluonic topological charge density 
to the nucleon, 
and $F$ a  renormalization scale invariant singlet decay constant; 
$g_{GNN}$ carries the renormalization  scale dependence of $g_A^{(0)}$.

\section{Glue in the $\eta'$}

While the pseudoscalar pions and kaons fit well as would-be Goldstone bosons 
with their masses satisfying $m_P^2 \propto m_q$, 
the isosinglet $\eta$ and 
$\eta'$ are too heavy by about 400 MeV and 300 MeV to satisfy this relation. 
They are exceptional mesons with their masses and interactions sensitive to additional non-perturbative gluon dynamics 
in the flavour-singlet channel~\cite{Fritzsch:1975tx,Crewther:1978zz,tHooft:1986ooh,DiVecchia:1980yfw,Witten:1980sp,Leutwyler:1997yr} 
associated 
with the QCD axial anomaly 
in the flavour-singlet axial-vector current. 
Within the approximation 
of a leading-order one-mixing-angle scheme, if we include a
${\tilde m}_{\eta_0}^2 =
0.73$ GeV$^2$  singlet gluonic mass term, then the $\eta$ and $\eta'$ masses each come out correct to within 10\% accuracy with the mixing angle
-20 degrees. 
Without this glue term in the isoscalar mass matrix, 
the $\eta'$ would come out as a strange quark state with mass 
$\sqrt{2 m_K^2 - m_\pi^2}$ 
and the $\eta$ would be a light quark state degenerate with the pion. 
The glue associated with ${\tilde m}_{\eta_0}^2$ 
is associated with gluon topology~\cite{Crewther:1978zz,tHooft:1986ooh} 
and 
its effect incorporated in axial U(1) extended chiral Lagrangians~\cite{DiVecchia:1980yfw,Witten:1980sp,Leutwyler:1997yr}. 
The theory involves the  interface of local anomalous Ward identities and 
non-local topological structure.
The $\eta$ and $\eta'$
masses then satisfy the Witten-Veneziano  mass formula 
$m_\eta^2 + m_{\eta'}^2 = 2m_K^2 + {\tilde m}_{\eta_0}^2$.
Recent lattice calculations 
with dynamical quarks of the meson masses and the gluonic term 
${\tilde m}^2_{\eta_0}$
further confirm this picture at 10\% accuracy~\cite{Cichy:2015jra}.
With the leading order mixing angle -20 degrees,  
the $\eta'$ has the biggest singlet component 
and hence the bigger sensitivity to OZI violating couplings to other hadrons 
proceeding through gluonic intermediate states. 
One expects OZI violation in the coupling of $\eta'$ to other hadrons in scattering processes over a broad range of energies~\cite{Bass:2018xmz}. 
This includes the 
low energy $\eta'$-nucleon coupling constant 
$g_{\eta' NN}$~\cite{Bass:1999is} 
through to large branching ratios observed in high energy 
$B \to \eta' X$ decays~\cite{Fritzsch:1997ps}.

Recent investigations have focussed on 
the properties of $\eta'$ mesons in nuclei. 
The light up and down quark contributions to the $\eta'$ wavefunction are induced by the gluonic term in the $\eta - \eta'$ mass matrix, and it is these that 
couple to the $\sigma$ 
(correlated two-pion) 
mean field inside the nucleus. 
Working within the Quark Meson Coupling model, QMC~\cite{Guichon:1987jp,Guichon:1995ue},  
a  
-37 MeV shift was predicted for the $\eta'$ mass for an 
$\eta'$ in a nucleus at nuclear matter density $\rho_0$ 
when the 
$\eta-\eta'$ mixing angle is taken as 
$\theta = -20$  degrees~\cite{Bass:2005hn,Bass:2013nya}.
Without the anomalous glue component in the $\eta'$ mass the $\eta'$ would be a strange quark state with much reduced interaction with the $\sigma$ mean field in the nucleus. 
$\eta'$ photoproduction experiments at ELSA in Bonn 
from Carbon and Niobium targets subsequently revealed an 
$\approx -40$ MeV shift in the $\eta'$ mass 
at $\rho_0$ with a small $\eta'$ width in medium.
The measured $\eta'$-nucleus
optical potential has real and imaginary parts $V+iW$ with 
\begin{eqnarray}
 V(\rho=\rho_0) &=& -40 \pm 6 \pm 15 \ {\rm MeV} 
\nonumber \\ 
W(\rho=\rho_0) &=& -13 \pm 3 \pm 3 \ {\rm MeV}   
\label{eq13}
\end{eqnarray}
-- see Refs.~\cite{CBELSATAPS:2013waf,Metag:2017yuh,Bass:2021rch}.
In these photoproduction experiments the meson is produced with reduced mass in the nucleus meaning that the production cross section goes up. 
When it emerges from the nucleus it returns to its free mass at expense of the kinetic energy. 
Data with a Carbon target shows
that the nuclear medium is approximately  transparent to $\eta'$ propagation, in contrast to  $\pi^0$  and $\eta$ propagation where there is large interaction with the nucleus~\cite{CBELSATAPS:2012few}.
The small $\eta'$ width relative to the mass shift (or potential depth) 
observed in these experiments 
means that 
possible $\eta'$ bound states in nuclei might be accessible in experiments. 
If observed, these bound states 
would be a new state of matter 
bound just by the strong interaction, in contrast to 
pionic and kaonic atoms 
involving electrically charged pions and kaons bound by QED interactions.
The present $\eta'$ bound state search experiments  
constrain the  possible parameter range 
with deep  potential depths of -150 MeV so far excluded~\cite{Eta-PRiMESuper-FRS:2016vbn,e-PRiMESuper-FRS:2017bzq,LEPS2BGOegg:2020cth}.  Further experiments in Germany and Japan  are running or in planning to push the these measurements towards  potentials typical of those suggested by the ELSA result and by QMC theory~\cite{Bass:2021rch,Saito:2023fnx}.

\section{The $\Delta$ excitation in polarized photoproduction from light nuclei}

A second interesting  nuclear effect 
involves 
the $\Delta$ resonance excitation 
in polarized photoproduction from 
polarized light nuclei where medium modification effects also enter.

The Gerasimov-Drell-Hearn, GDH, sum-rule for polarized photoproduction relates
the inclusive 
spin cross sections for polarized photon-proton scattering 
to the ratio of the target proton anomalous magnetic moment and mass.
The GDH sum-rule reads\cite{Gerasimov:1965et,Drell:1966jv} 
\begin{equation}
\int^{\infty}_{M^2} 
\frac{ds_{\gamma p}}{s_{\gamma p} - M^2}
(\sigma_{\rm p} - \sigma_{\rm a})
= 
\frac{4 S \pi^2 \alpha 
\kappa^2}{M^2}
\label{eq14}
\end{equation}
where $\sigma_{\rm p}$ and $\sigma_{\rm a}$ 
are the spin dependent photoabsorption cross-sections
involving photons polarized parallel and antiparallel to the target's spin.
Here
$s_{\gamma p}$ 
is the photon-proton centre-of-mass energy squared
with $\kappa$ the target's anomalous magnetic moment;
$M$ is the target mass and $S=\frac{1}{2}$ is its spin.
The GDH 
sum-rule is derived from the very general principles of causality, unitarity,
Lorentz and electromagnetic gauge invariance
together with the single assumption that 
$\sigma_{\rm p} - \sigma_{\rm a}$
satisfies an unsubtracted dispersion relation.

For free protons 
the GDH sum-rule predicts a value of $+205 \ \mu$b
for the integral in Eq.~(\ref{eq14}) with the proton's anomalous magnetic moment $\kappa=1.79$.
This result is in excellent agreement with the experimental number 
$+209 \pm 13$ $\mu$b.
This value is extracted as follows. 
Polarized photoproduction  experiments at MAMI in Mainz and ELSA in Bonn 
have measured 
$\sigma_{\rm p} - \sigma_{\rm a}$ 
with photon beam energies between 200 MeV and 2.9 GeV giving a measured 
sum-rule contribution
$+253.5 \pm 5 \pm 12$ $\mu$b~\cite{Dutz:2004zz,Helbing:2006zp}.
Here the biggest contribution comes from the 
$\Delta$ 
magnetic transition excitation,  
about 190 $\mu$b, with smaller contributions from heavier resonances.
One also has a 
near threshold contribution $\approx -30$ $\mu$b estimated 
from multipole models of pion photoproduction~\cite{Drechsel:2004ki}.
A further $\approx 10$\% part of the sum-rule comes 
from high energy  Regge contributions from energies above the maximum ELSA beam energy, 
$-15 \pm 2$ $\mu$b \cite{Bass:2018uon},  
estimated from Regge fits to low $Q^2$ leptoproduction data. 
The high energy 
part of the proton GDH sum-rule is  
essentially all in the isovector channel with negligible isoscalar contribution,
similar to the situation observed with  polarized deep inelastic scattering in the so far  measured kinematics -- see Fig.~\ref{fig:g1int}.
An independent confirmation of the GDH sum-rule comes from JLab,  
$+204 \pm 11$ $\mu$b 
from extrapolation 
of low $Q^2$ electroproduction  data to the real photon point~\cite{CLAS:2021apd}.

How about possible changes of 
nucleon properties in polarized photoproduction from (light) nuclei?
Extending these GDH experiments 
from protons to 
light nuclear targets, 
experiments 
have so far been performed also with polarized deuterons and $^3$He targets. 
The present data involves measurements with photon energies from 200 MeV up to 1.8 GeV (deuteron data)~\cite{Ahrens:2009zz} and up to 500 MeV ($^3$He data)~\cite{AguarBartolome:2013mga}.

As recently noticed~\cite{Bass:2022pyx}, 
these data suggest a 
small shift in 
the excitation energy of the $\Delta$ peak, up to  
$\approx -20$ MeV,
in the spin difference cross section 
$\sigma_{\rm p} - \sigma_{\rm a}$ 
(specifically in the spin
parallel cross section $\sigma_{\rm p}$ and not observed in the spin average cross section) 
with the effect most visible in the more precise deuteron data. 
This observed cross sections are 
qualitatively different from what one expects from smearing due to Fermi motion of the bound nucleons in the light nuclei. 
Small, few percent,  medium modifications of nucleon properties 
in light deuterons 
are also observed in experimental measurements of the EMC nuclear effect 
where parton distributions of bound nucleons in the deuteron are seen 
to be modified relative to free protons~\cite{Griffioen:2015hxa}, 
as well as with theoretical lattice calculations of the 
nucleon's axial and tensor charges in light nuclei~\cite{Chang:2017eiq}.

It would be interesting to investigate whether 
the $\Delta$ peak mass shift effect persists and might be enhanced 
with a larger 
polarized nuclear target where medium effects might be more pronounced.
Experimentally, one needs 
a target where the spin of nucleus is carried close to all by a single polarized nucleon, 
e.g. $^7$Li
where the nucleus is not so big that any effect is washed out by a 
huge spin dilution factor in the total asymmetry from scattering on unpolarized spectator nucleons carrying close to no net spin. 
More generally, if one could measure the 
GDH sum-rule for a bound nucleon over the full energy range,  
then the right-hand side 
static part would become sensitive to medium modifications in the proton mass and anomalous magnetic moment.  
Model estimates of these quantities suggest a possible enhancement in the sum-rule up to about a factor of two for a polarized proton at nuclear matter density \cite{Bass:2022pyx,Bass:2020bkl}.

\section{Conclusions}

QCD continues to inspire new advances in theory and experiments in our quest to understand the deep structure of the proton. 
In the topics of proton spin dynamics and $\eta'$ interactions new data will soon follow 
from near threshold production processes up to high energy deep inelastic scattering. 
The physics puzzles that inspired Harald Fritzsch continue to inspire us today.

\bibliographystyle{ws-rv-van}
\bibliography{bass-bib}

\begin{thebibliography}{10}
\expandafter\ifx\csname urlstyle\endcsname\relax
  \providecommand{\doi}[1]{doi:\discretionary{}{}{}#1}\else
  \providecommand{\doi}{doi:\discretionary{}{}{}\begingroup \urlstyle{rm}\Url}\fi

\bibitem{Fritzsch:1973pi}
H.~Fritzsch, M.~Gell-Mann and H.~Leutwyler, {\em Phys. Lett. B} {\bf 47}, 365  (1973).

\bibitem{Fritzsch:1972jv}
H.~Fritzsch and M.~Gell-Mann, {\it {Current algebra: Quarks and what else?}}, in {\em Proceedings of 16th International Conference on High-Energy Physics, Batavia, Illinois, 6-13 Sept.\/},  eds. J.~D. Jackson and A.~Roberts, {\em eConf} {\bf C720906V2} (1972).
\newblock pp. 135--165.
\newblock \url{hep-ph/0208010}.

\bibitem{Gross:1973id}
D.~J. Gross and F.~Wilczek, {\em Phys. Rev. Lett.} {\bf 30}, 1343  (1973).

\bibitem{Politzer:1973fx}
H.~D. Politzer, {\em Phys. Rev. Lett.} {\bf 30}, 1346  (1973).

\bibitem{Close:1979bt}
F.~E. Close, {\em {An Introduction to Quarks and Partons}} (Academic N.Y., 1979).

\bibitem{Gasser:1982ap}
J.~Gasser and H.~Leutwyler, {\em Phys. Rept.} {\bf 87}, 77  (1982).

\bibitem{Collins:1976yq}
J.~C. Collins, A.~Duncan and S.~D. Joglekar, {\em Phys. Rev. D} {\bf 16}, 438  (1977).

\bibitem{Shifman:1988zk}
M.~A. Shifman, {\em Phys. Rept.} {\bf 209}, 341  (1991).

\bibitem{Jaffe:1989jz}
R.~L. Jaffe and A.~Manohar, {\em Nucl. Phys. B} {\bf 337}, 509  (1990).

\bibitem{Nambu:1961tp}
Y.~Nambu and G.~Jona-Lasinio, {\em Phys. Rev.} {\bf 122}, 345  (1961).

\bibitem{Klevansky:1992qe}
S.~P. Klevansky, {\em Rev. Mod. Phys.} {\bf 64}, 649  (1992).

\bibitem{Larin:1997qq}
S.~A. Larin, T.~van Ritbergen and J.~A.~M. Vermaseren, {\em Phys. Lett. B} {\bf 404}, 153  (1997).

\bibitem{Blumlein:2022gpp}
J.~Bl\"umlein, P.~Marquard, C.~Schneider and K.~Sch\"onwald, {\em JHEP} {\bf 11},   156  (2022).

\bibitem{Bass:2004xa}
S.~D. Bass, {\em Rev. Mod. Phys.} {\bf 77}, 1257  (2005).

\bibitem{Workman:2022ynf}
Particle Data Group Collaboration, R.~L. Workman {\em et~al.}, {\em PTEP} {\bf 2022},   083C01  (2022).

\bibitem{Close:1993mv}
F.~E. Close and R.~G. Roberts, {\em Phys. Lett. B} {\bf 316}, 165  (1993).

\bibitem{COMPASS:2016jwv}
COMPASS Collaboration, C.~Adolph {\em et~al.}, {\em Phys. Lett. B} {\bf 769}, 34  (2017).

\bibitem{EuropeanMuon:1987isl}
European Muon Collaboration, J.~Ashman {\em et~al.}, {\em Phys. Lett. B} {\bf 206},   364  (1988).

\bibitem{Landshoff:1994up}
P.~V. Landshoff  (1994), \url{hep-ph/9410250}.

\bibitem{Aidala:2012mv}
C.~A. Aidala, S.~D. Bass, D.~Hasch and G.~K. Mallot, {\em Rev. Mod. Phys.} {\bf 85}, 655  (2013).

\bibitem{Altarelli:1988nr}
G.~Altarelli and G.~G. Ross, {\em Phys. Lett. B} {\bf 212}, 391  (1988).

\bibitem{Efremov:1988zh}
A.~V. Efremov and O.~V. Teryaev (1988).
\newblock JINR-E2-88-287.

\bibitem{Carlitz:1988ab}
R.~D. Carlitz, J.~C. Collins and A.~H. Mueller, {\em Phys. Lett. B} {\bf 214}, 229  (1988).

\bibitem{Bass:1991yx}
S.~D. Bass, B.~L. Ioffe, N.~N. Nikolaev and A.~W. Thomas, {\em J. Moscow. Phys. Soc.} {\bf 1}, 317  (1991).

\bibitem{Bass:1997zz}
S.~D. Bass, {\em Mod. Phys. Lett. A} {\bf 13}, 791  (1998).

\bibitem{Bass:2002mv}
S.~D. Bass, R.~J. Crewther, F.~M. Steffens and A.~W. Thomas, {\em Phys. Rev. D} {\bf 66},   031901  (2002).

\bibitem{Broadhurst:1973fr}
D.~J. Broadhurst, J.~F. Gunion and R.~L. Jaffe, {\em Annals Phys.} {\bf 81},  ~88  (1973).

\bibitem{Jaffe:1995an}
R.~L. Jaffe, {\em Phys. Lett. B} {\bf 365}, 359  (1996).

\bibitem{Manohar:1990kr}
A.~V. Manohar, {\em Phys. Rev. Lett.} {\bf 65}, 2511  (1990).

\bibitem{Bass:1992ti}
S.~D. Bass, {\em Z. Phys. C} {\bf 55}, 653  (1992).

\bibitem{deFlorian:2014yva}
D.~de~Florian, R.~Sassot, M.~Stratmann and W.~Vogelsang, {\em Phys. Rev. Lett.} {\bf 113},   012001  (2014).

\bibitem{Bass:2011zn}
S.~D. Bass, A.~Casey and A.~W. Thomas, {\em Phys. Rev. C} {\bf 83},   038202  (2011).

\bibitem{Brodsky:1989db}
S.~J. Brodsky and I.~A. Schmidt, {\em Phys. Lett. B} {\bf 234}, 144  (1990).

\bibitem{deFlorian:2019egz}
D.~de~Florian and W.~Vogelsang, {\em Phys. Rev. D} {\bf 99},   054001  (2019).

\bibitem{Zhou:2022wzm}
JAM Collaboration, Y.~Zhou, N.~Sato and W.~Melnitchouk, {\em Phys. Rev. D} {\bf 105},   074022  (2022).

\bibitem{Bass:2009ed}
S.~D. Bass and A.~W. Thomas, {\em Phys. Lett. B} {\bf 684}, 216  (2010).

\bibitem{Ethier:2017zbq}
J.~J. Ethier, N.~Sato and W.~Melnitchouk, {\em Phys. Rev. Lett.} {\bf 119},   132001  (2017).

\bibitem{Alexandrou:2017oeh}
C.~Alexandrou {\em et~al.}, {\em Phys. Rev. Lett.} {\bf 119},   142002  (2017).

\bibitem{Liang:2018pis}
J.~Liang, Y.-B. Yang, T.~Draper, M.~Gong and K.-F. Liu, {\em Phys. Rev. D} {\bf 98},   074505  (2018).

\bibitem{Adamiak:2023okq}
D.~Adamiak, Y.~V. Kovchegov and Y.~Tawabutr, {\em Phys. Rev. D} {\bf 108},   054005  (2023).

\bibitem{Adamiak:2021ppq}
JAM Collaboration, D.~Adamiak {\em et~al.}, {\em Phys. Rev. D} {\bf 104},   L031501  (2021).

\bibitem{KamLAND:2022ptk}
KamLAND Collaboration, S.~Abe {\em et~al.}, {\em Phys. Rev. D} {\bf 107},   072006  (2023).

\bibitem{Ji:1996ek}
X.-D. Ji, {\em Phys. Rev. Lett.} {\bf 78}, 610  (1997).

\bibitem{Bhattacharya:2022xxw}
S.~Bhattacharya, Y.~Hatta and W.~Vogelsang, {\em Phys. Rev. D} {\bf 107},   014026  (2023).

\bibitem{Bhattacharya:2023wvy}
S.~Bhattacharya, Y.~Hatta and W.~Vogelsang, {\em Phys. Rev. D} {\bf 108},   014029  (2023).

\bibitem{Bass:2001dg}
S.~D. Bass, {\em Phys. Rev. D} {\bf 65},   074025  (2002).

\bibitem{Aidala:2021pvc}
C.~A. Aidala and T.~C. Rogers, {\em Phil. Trans. A. Math. Phys. Eng. Sci.} {\bf 380},   20210058  (2021).

\bibitem{Kharzeev:2021nzh}
D.~E. Kharzeev, {\em Phil. Trans. A. Math. Phys. Eng. Sci.} {\bf 380},   20210063  (2021).

\bibitem{Fritzsch:1988ht}
H.~Fritzsch, {\em Mod. Phys. Lett. A} {\bf 5},   625  (1990).

\bibitem{Fritzsch:1989ty}
H.~Fritzsch, {\em Phys. Lett. B} {\bf 229}, 122  (1989).

\bibitem{Forte:1990xb}
S.~Forte and E.~V. Shuryak, {\em Nucl. Phys. B} {\bf 357}, 153  (1991).

\bibitem{Shore:1991dv}
G.~M. Shore and G.~Veneziano, {\em Nucl. Phys. B} {\bf 381}, 23  (1992).

\bibitem{Tarasov:2020cwl}
A.~Tarasov and R.~Venugopalan, {\em Phys. Rev. D} {\bf 102},   114022  (2020).

\bibitem{Tarasov:2021yll}
A.~Tarasov and R.~Venugopalan, {\em Phys. Rev. D} {\bf 105},   014020  (2022).

\bibitem{Veneziano:1989ei}
G.~Veneziano, {\em Mod. Phys. Lett. A} {\bf 4},   1605  (1989).

\bibitem{Bass:1991sg}
S.~D. Bass, {\em Int. J. Mod. Phys. A} {\bf 7}, 6039  (1992).

\bibitem{Fritzsch:1975tx}
H.~Fritzsch and P.~Minkowski, {\em Nuovo Cim. A} {\bf 30},   393  (1975).

\bibitem{Crewther:1978zz}
R.~J. Crewther, {\em Acta Phys. Austriaca Suppl.} {\bf 19}, 47  (1978).

\bibitem{tHooft:1986ooh}
G.~'t~Hooft, {\em Phys. Rept.} {\bf 142}, 357  (1986).

\bibitem{DiVecchia:1980yfw}
P.~Di~Vecchia and G.~Veneziano, {\em Nucl. Phys. B} {\bf 171}, 253  (1980).

\bibitem{Witten:1980sp}
E.~Witten, {\em Annals Phys.} {\bf 128},   363  (1980).

\bibitem{Leutwyler:1997yr}
H.~Leutwyler, {\em Nucl. Phys. B Proc. Suppl.} {\bf 64}, 223  (1998).

\bibitem{Cichy:2015jra}
ETM Collaboration, K.~Cichy {\em et~al.}, {\em JHEP} {\bf 09},   020  (2015).

\bibitem{Bass:2018xmz}
S.~D. Bass and P.~Moskal, {\em Rev. Mod. Phys.} {\bf 91},   015003  (2019).

\bibitem{Bass:1999is}
S.~D. Bass, {\em Phys. Lett. B} {\bf 463}, 286  (1999).

\bibitem{Fritzsch:1997ps}
H.~Fritzsch, {\em Phys. Lett. B} {\bf 415}, 83  (1997).

\bibitem{Guichon:1987jp}
P.~A.~M. Guichon, {\em Phys. Lett. B} {\bf 200}, 235  (1988).

\bibitem{Guichon:1995ue}
P.~A.~M. Guichon, K.~Saito, E.~N. Rodionov and A.~W. Thomas, {\em Nucl. Phys. A} {\bf 601}, 349  (1996).

\bibitem{Bass:2005hn}
S.~D. Bass and A.~W. Thomas, {\em Phys. Lett. B} {\bf 634}, 368  (2006).

\bibitem{Bass:2013nya}
S.~D. Bass and A.~W. Thomas, {\em Acta Phys. Polon. B} {\bf 45},   627  (2014).

\bibitem{CBELSATAPS:2013waf}
CBELSA/TAPS Collaboration, M.~Nanova {\em et~al.}, {\em Phys. Lett. B} {\bf 727}, 417  (2013).

\bibitem{Metag:2017yuh}
V.~Metag, M.~Nanova and E.~Y. Paryev, {\em Prog. Part. Nucl. Phys.} {\bf 97}, 199  (2017).

\bibitem{Bass:2021rch}
S.~D. Bass, V.~Metag and P.~Moskal  (2022), in: Tanihata, I., Toki, H., Kajino, T. (eds) Handbook of Nuclear Physics. Springer, Singapore. \url{arXiv:2111.01388 [hep-ph]}.

\bibitem{CBELSATAPS:2012few}
CBELSA/TAPS Collaboration, M.~Nanova {\em et~al.}, {\em Phys. Lett. B} {\bf 710}, 600  (2012).

\bibitem{Eta-PRiMESuper-FRS:2016vbn}
$\eta$-PRiME/Super-FRS Collaboration, Y.~K. Tanaka {\em et~al.}, {\em Phys. Rev. Lett.} {\bf 117},   202501  (2016).

\bibitem{e-PRiMESuper-FRS:2017bzq}
\ensuremath{\eta}-PRiME/Super-FRS Collaboration, Y.~K. Tanaka {\em et~al.}, {\em Phys. Rev. C} {\bf 97},   015202  (2018).

\bibitem{LEPS2BGOegg:2020cth}
LEPS2/BGOegg Collaboration, N.~Tomida {\em et~al.}, {\em Phys. Rev. Lett.} {\bf 124},   202501  (2020).

\bibitem{Saito:2023fnx}
T.~R. Saito {\em et~al.}, {\em Nucl. Instrum. Meth. B} {\bf 542}, 22  (2023).

\bibitem{Gerasimov:1965et}
S.~B. Gerasimov, {\em Yad. Fiz.} {\bf 2}, 598  (1965).

\bibitem{Drell:1966jv}
S.~D. Drell and A.~C. Hearn, {\em Phys. Rev. Lett.} {\bf 16}, 908  (1966).

\bibitem{Dutz:2004zz}
H.~Dutz {\em et~al.}, {\em Phys. Rev. Lett.} {\bf 93},   032003  (2004).

\bibitem{Helbing:2006zp}
K.~Helbing, {\em Prog. Part. Nucl. Phys.} {\bf 57}, 405  (2006).

\bibitem{Drechsel:2004ki}
D.~Drechsel and L.~Tiator, {\em Ann. Rev. Nucl. Part. Sci.} {\bf 54}, 69  (2004).

\bibitem{Bass:2018uon}
S.~D. Bass, M.~Skurzok and P.~Moskal, {\em Phys. Rev. C} {\bf 98},   025209  (2018).

\bibitem{CLAS:2021apd}
CLAS Collaboration, X.~Zheng {\em et~al.}, {\em Nature Phys.} {\bf 17}, 736  (2021).

\bibitem{Ahrens:2009zz}
J.~Ahrens {\em et~al.}, {\em Phys. Lett. B} {\bf 672}, 328  (2009).

\bibitem{AguarBartolome:2013mga}
P.~Aguar~Bartolome {\em et~al.}, {\em Phys. Lett. B} {\bf 723}, 71  (2013).

\bibitem{Bass:2022pyx}
S.~D. Bass, P.~Pedroni and A.~Thomas, {\em Eur. Phys. J. A} {\bf 59},   239  (2023).

\bibitem{Griffioen:2015hxa}
K.~A. Griffioen {\em et~al.}, {\em Phys. Rev. C} {\bf 92},   015211  (2015).

\bibitem{Chang:2017eiq}
NPLQCD Collaboration, E.~Chang {\em et~al.}, {\em Phys. Rev. Lett.} {\bf 120},   152002  (2018).

\bibitem{Bass:2020bkl}
S.~D. Bass, {\em Acta Phys. Polon. B} {\bf 52}, 43  (2021).

\end{thebibliography}

\end{document}